\documentclass[aps,prb]{revtex4-2}

\usepackage{bm}
\usepackage{amsmath}
\usepackage{color}
\usepackage{graphicx}
\usepackage{amssymb}

\DeclareMathAlphabet{\mathbbmsl}{U}{bbm}{m}{sl}
\makeatletter
\newsavebox{\@brx}

\begin{document}

\title{Polarizability, plasmons, and screening in  1T$^\prime$-MoS$_2$ with  tilted Dirac bands}

\def\correspondingauthor{\footnote{Corresponding author: balassis@fordham.edu}}

\author{Antonios Balassis$^{1}$\correspondingauthor{},  Godfrey Gumbs$^{2}$, and Oleksiy Roslyak$^{1}$ }

\affiliation{$^1$ Department of Physics \& Engineering Physics, Fordham University,  441 East Fordham Road,  Bronx, NY  10458 USA}
\affiliation{$^{2}$Department of Physics and Astronomy, Hunter College of the
City University of New York, 695 Park Avenue, New York, NY 10065, USA}

\date{\today}

\begin{abstract}
   In the presence of an external vertical electric field and strain, it is evident that 1T$^\prime$-MoS$_2$  exhibits tilted Dirac bands  which are valley-spin-polarized.  Additionally, this material experiences  a topological phase change between a topological insulator and band insulator  for a critical value of the electric field.  Using linear response theory, we calculated the polarization function  which is in turn employed to obtain the dielectric function. This latter quantity is subsequently utilized in calculations to determine the plasmons dispersion relation, their decay rate and impurity screening  corresponding to various levels of doping, the critical applied vertical electric field strengths and the spin-orbit coupling gap in 1T$^\prime$-MoS$_2$ with tilted Dirac bands.  
\end{abstract}

\maketitle

\section{Introduction}
\label{sec1}

Massive anisotropic tilted Dirac systems have been gaining an increasing degree of attention  among two-dimensional (2D)   materials due to their intriguing properties.\cite{Science,first,second,third,fourth,fifth} These include their distinctive anisotropic optical response. \cite{first} Materials like graphene and silicene  have an isotropic  relativistic spectrum in momentum space. \cite{Advan,Rev2}  However,   it has been found that some of these new materials under consideration have anisotropic  linear spectra, namely, tilted anisotropic linear Dirac cones, which is the case for  1T$^\prime$-MoS$_2$  \cite{Science,first} and 8-{\em Pmmn} borophene, \cite{15,16,17,18} partially hydrogenated graphene \cite{26} as well as    transition metal dichalcogenides. \cite{first,24} In effect, the existence of a tilted cone gives rise to fundamentally different electronic and optical behaviors compared with systems whose cones are not tilted.\  \cite{27} For example, anisotropic plasmon dispersion was reported in Ref.  \cite{28}  as well as a unique intervalley damping effect.  \cite{29,30}

\medskip
\par
These materials are semimetals and have no intrinsic band gap, but a gap can be generated by breaking the symmetry, for example.  In the case of graphene,  the semimetalic behavior can be changed by breaking the inversion symmetry and the opening of a band gap, as demonstrated in Refs.\  \cite{AI1,AI2}, and which leads to the valley Hall effect. It has been found that in the presence of external vertical electric field, 1T$^\prime$-MoS$_2$ presents valley-spin-polarized tilted Dirac bands.  Additionally, for a critical value of the electric field,  the system undergoes a topological transition between the topological insulator and band insulator phases.  We investigate the effects due to the vertical electric field and doping  on the anisotropic   polarization function, plasmon excitations and their decay rates  as well as impurity screening   for tilted Dirac bands for 1T$^\prime$-MoS$_2$ at T=0 K.  These calculations, done using linear response theory, reveal the role played by the combined effect due to spin-orbit coupling, band tilting, and vertical electric field on an important collective property with potential device applications.    Our results for 1T$^\prime$-MoS$_2$ are compared with   8-{\em Pmmn} borophene, a polymorph of borophene which has an anisotropic tilted Dirac cone \cite{28}, and other monolayer tilted gapped Dirac materials, including  $\alpha-$SnS$_2$, TaCoTe$_2$, and TaIrTe$_4$, based on the similarity of their band structure.

\medskip
\par
The rest of this paper is organized as follows.    For our theory, we introduce in Sec.\  \ref{sec2} a low-energy model Hamiltonian for 1T$^\prime$-MoS$_2$. We analytically derive the eigenstates.    In Sec.\  \ref{sec3}, we derive the frequency-dependent polarization function   and present our numerical results.   Section \ref{sec4}  is devoted to an investigation of the plasmon excitations and their decay rates for chosen doping and vertical electric field strength.  The static shielding of a dilute distribution of impurities   is presented in Sec.\  \ref{sec5}. We conclude with a summary in Sec.\  \ref{sec6}.

\section{Anisotropic Tilted Dirac Bands Formalism}
 \label{sec2}

 Within the ${\mathbf k} \cdot {\mathbf p}$ approximation the low-energy Hamiltonian for a 2D anisotropic tilted Dirac system representing 1T$^\prime$ -MoS$_2$ in the vicinity of two independent Dirac points  located at (0, $\lambda\Lambda$) with $\lambda=\pm1$ is given by \cite{first}

\begin{eqnarray}
\hat{{\cal H}}_\lambda(k_x,k_y) &=& \hbar k_x v_1\gamma_1 + \hbar k_y ( v_2\gamma_0 -\lambda v_- {\bf I} -\lambda v_+\gamma_2)
\nonumber\\
&+& \Delta  (\lambda \gamma_0-i\alpha\gamma_1\gamma_2) \  .
\label{eq1}
\end{eqnarray}
In this notation,  the 2D wave vector ${\bf k}=(k_x,k_y)$, the spin-orbit coupling  parameter
$\Delta =0.042$ eV and  the Fermi velocities are given by $v_1=3.87\times 10^5$ m/s, $v_2=0.46\times 10^5$ m/s, $v_-=2.86\times 10^5$ m/s, and $v_+=7.21\times 10^5$ m/s.  Also, ${\bf I}$ is the unit $4\times 4$ matrix and we have introduced $\gamma_0=\tau_1\otimes \sigma_1$, $\gamma_1=\tau_2\otimes \sigma_0$,  and $\gamma_2=\tau_3\otimes \sigma_0$ with, $\tau_0$ and $\tau_i$ being unit and Pauli matrices acting in pseudospin space whereas $\gamma_0$ and $\gamma_i$ denote Pauli matrices acting upon real spin space.  The normalized electric field $\alpha=|E_z|/|E_c|$ is defined via the applied electric field $E_z$ and its critical value $E_c$.

\medskip
\par

\begin{figure}[ht]
\centering
\includegraphics[width=0.8\textwidth]{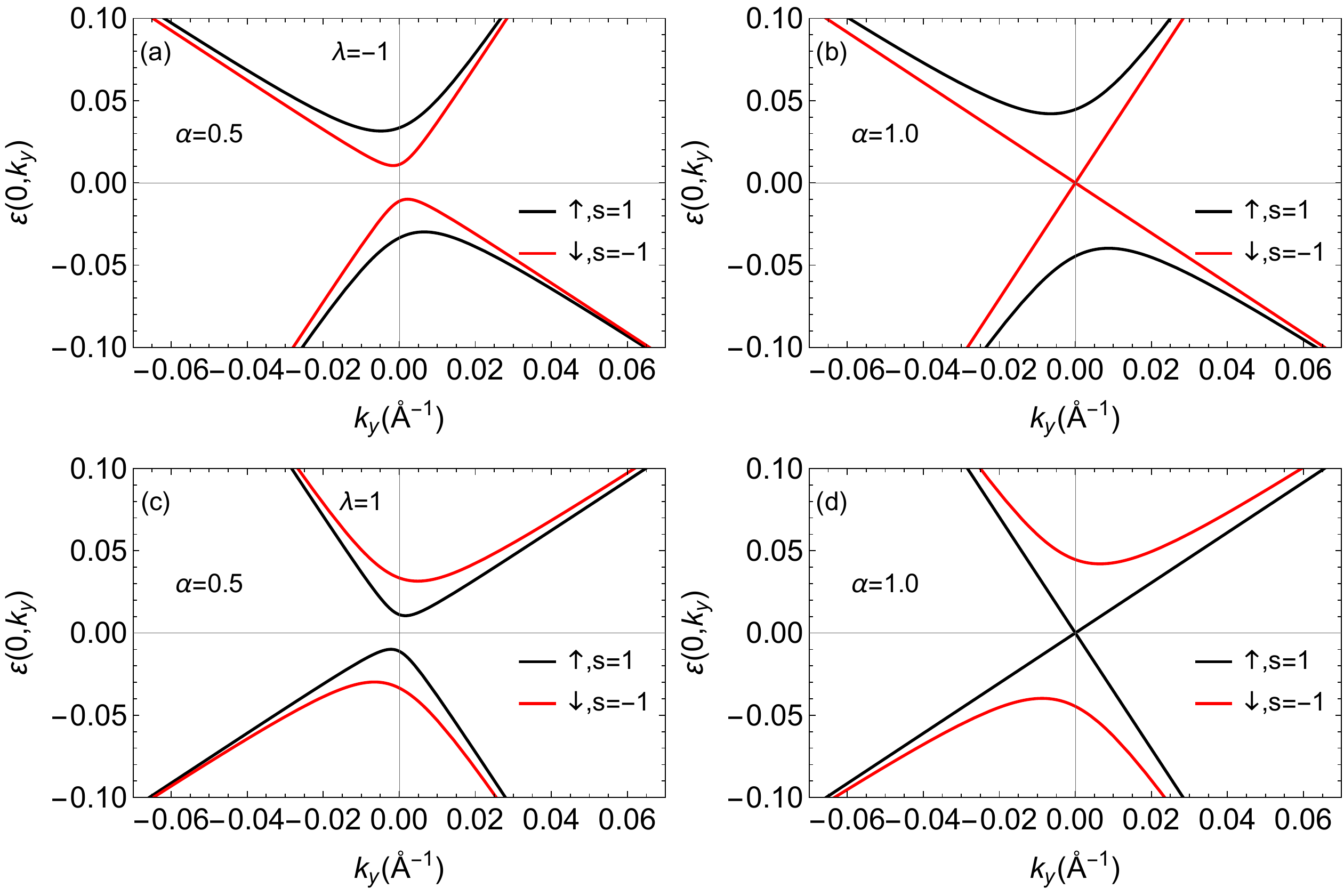}
\caption{ (Color online)   Anisotropic tilted energy bands at the (0, $\lambda$)  points  for $\alpha=0.5$ and $\alpha=1.0$ when $k_x=0$. The unit for the energy, is $\hbar k_{y} v_{-}=1.88$ eV ($k_{y}=1\textrm{\AA}^{-1}$).  Panels (a)  and (b) are for $\lambda=-1$, panels (c) and (d) are for $\lambda=1$. }
\label{Fig1}
\end{figure}

The energy eigenvalues of  Eq.~(\ref{eq1}) are given by

\begin{equation}
   \epsilon_{\xi,s}^{\lambda}({\bf k})=-\lambda \hbar v_{-} k_{y}+\xi \sqrt{\left[\hbar v_{2} k_{y}+\left(\lambda-s\alpha\right)\Delta  \right]^2+\left(\hbar v_{1} k_{x}\right)^2+\left(\hbar v_{+} k_{y}\right)^2} \  ,
\label{ANSWER}
\end{equation}
where $\xi=\pm 1$ for the conduction (valence) band, and $s=\pm 1$ is the spin up (down) index.
The long wavelength expansion of the energy eigenvalues for applied electric field not close to its critical value, $\alpha \neq 1$, is 

\begin{align}
  \epsilon_{\xi,s}^{\lambda}({\bf k})=  \xi |\lambda - s \alpha| \Delta+\left\{  -\lambda \hbar v_{-} +\xi \hbar v_{2}\  \text{sgn}(\lambda-s\alpha) \right\}k_{y}+\left(\dfrac{\xi \hbar^2 }{2\Delta|\lambda-\alpha s|}\right)\left(v_1^2 k_{x}^2+ v_{+}^2{k_y}^2\right),
\label{longwave}
\end{align}
and we see that it depends on both $k_y$ (a linear term) and $k_y^2$  but only on a $k_x^2$ quadratic term. Equation~(\ref{longwave}) also shows  that the spin-orbit coupling opens up a gap between spin-subbands  and between the valence and conduction bands within a chosen valley. We emphasize that Eq.~(\ref{longwave}) is not valid in the gapless case.

 \medskip
\par
In Fig.~\ref{Fig1}, we show the spin-polarized bands  and the valley-spin polarized gaps in the presence of a vertical electric field  ($\alpha\neq 0$). Notice that an indirect energy gap opens when  $\alpha \neq 1$.  The critical points  of the energy bands are located at ${\bf k}^\ast =(0, \kappa(s,\xi,\lambda))$  whose coordinate on the $k_y$-axis    for the minima (for $\xi=1$) or maxima (for $\xi=-1$) of Eq.~(\ref{ANSWER}) are

\begin{equation}
\kappa(s,\xi,\lambda)  =\frac{\left(\xi v_{+} v_{-}-v_2\sqrt{v_2^2-v_{-}^2+v_{+}^2}\right)\left(\lambda-s\alpha \right)\Delta}{\hbar \left(v_2^2+v_{+}^2\right)\sqrt{v_2^2-v_{-}^2+v_{+}^2}} \ .
\label{AO}
\end{equation}
According to Eq.\ (\ref{AO}), the critical point is at the origin when $\lambda=s\alpha$ which is in agreement with the panels on the right-hand side in Fig.\   \ref{Fig1}  for both spin branches and valleys.  We now have an explicit expression  for the shifted critical point as well as the tilt by taking the gradient of the expression in Eq.~(\ref{longwave}).
 
\medskip
\par 

The wave functions of  the Hamiltonian in  Eq.~(\ref{eq1}) are given by the following  expression

\begin{equation}
\psi_{\xi,s}^{\lambda}  ({\bf k},{\bf r})=\left(
\begin{array}{c}
\dfrac{-\lambda \hbar k_{y} v_{+}+\xi \sqrt{|G|^{2}+\left(\hbar k_{y} v_{+}\right)^{2}}}{G} \\[6pt]
\dfrac{\lambda \hbar k_{y} v_{+}-\xi \sqrt{|G|^{2}+\left(\hbar k_{y} v_{+}\right)^{2}}}{sG}) \\[6pt]
-s\\[6pt]
1
\end{array} \right)
\frac{e^{i{\bf k}\cdot {\bf r}}}{\sqrt{A}}, 
\end{equation}
where $A$ is a normalization area and we have introduced  $G\equiv  (\lambda-s \alpha) \Delta+\hbar v_{2} k_{y}-  i s \hbar v_{1} k_{x}$.

\section{Frequency-dependent Polarization function}
\label{sec3}

\medskip
\par 

In the RPA,  the polarizability  is represented by a particle-hole bubble in the in the Feynman diagram representation. Mathematically, this is for each channel

\begin{equation}
\Pi_{\lambda}  ({\bf q},\omega)=    \sum_{\xi,s}\sum_{\xi',s'}\int  \frac{d^2{\bf k}}{(2\pi)^2}
 \frac{f(\epsilon_{\xi',s'}^{\lambda}({\bf k}+{\bf q}))-f(\epsilon_{\xi,s}^{\lambda}({\bf k}))}{\hbar \omega+\epsilon_{\xi',s'}^{\lambda}({\bf k}+{\bf q})-\epsilon_{\xi,s}^{\lambda}({\bf k})+i \delta}F^{\lambda}_{\xi, s,\xi',s'}({\bf k},{\bf q}) \ ,
\label{pol1}
\end{equation}
where  $f(\epsilon_{\xi,s}^{\lambda}({\bf k}))$  is the Fermi-Dirac distribution function under chemical potential $\mu$  and $\epsilon_{\xi,s}^{\lambda}({\bf k})$  are the energy eigenvalues,
$\delta= 0^+$  can be viewed as an infinitesimal scattering rate. The numerator containing the two statistical functions ensures that the integrand covers only the overlap of particle-hole pairs     $|\lambda,\xi,s,{\bf k}>$   and $|\lambda,\xi^\prime,s^\prime,{\bf k}+{\bf q}>$ (as opposed to particle-particle or hole-hole pairs). The definition of the overlap function   $F^{\lambda}_{\xi,s,\xi',s'}({\bf k},{\bf q})$ is

\begin{equation}
F^{\lambda}_{\xi,s,\xi',s'}({\bf k},{\bf q})) \equiv |   \left<\psi_{\xi,s}^
\lambda({\bf k})|
\psi_{\xi',s'}^{\lambda }({\bf k}+{\bf q}) \right >|^2 \ .
\label{overlap}
\end{equation}

\medskip
\par 

\begin{figure}[h]
\centering
\includegraphics[width=1.0\textwidth]{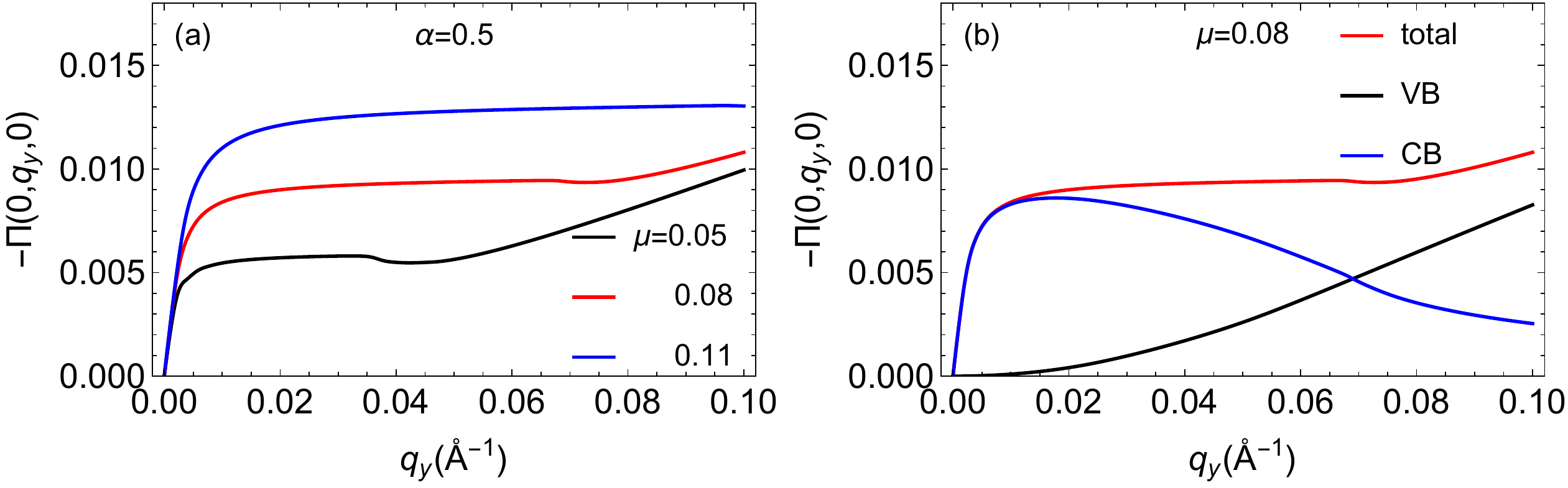}
\caption{ (Color online)  (a) Static  polarization using the $\lambda=\pm 1$ energy bands of Fig.~\ref{Fig1} for $\alpha=0.5$ and various values of the chemical potential $\mu$ in the conduction band. The chemical potential has the same unit as the energy, (b) Contributions from the valence band (VB) and from the conduction band (CB) to the total polarization for $\mu=0.083$.}
\label{Fig2}
\end{figure}

The parameters used  in our calculations are the same as before. In Fig.~\ref{Fig2},  we present the static polarization function versus wave number at zero temperature for various chemical potentials, chosen $\alpha=0.5$ and $1.0$. We have included the contributions to the polarizability from both   valleys, i.e., $\lambda=\pm 1$.  More explicitly, in  Figs.~\ref{Fig2}(a) and \ref{Fig2}(b),  we set $\lambda=\pm 1$ in the sum in Eq.~ (\ref{pol1}) and compare the total value of the polarization function as the chemical potential is varied in Fig.~\ref{Fig2}(a) whereas Fig.~\ref{Fig2}(b) shows the individual contributions from the valence (VB) and the conduction (CB) to the total polarizability. These results show that the value of the polarization is increased as the chemical potential is increased but its variation  with $q_y$ is not monotonic.  Figure \ref{Fig2}(b)  shows that the contribution to the polarizability  is dominated by the conduction band (CB) at longer wavelengths but the valence band (VB) makes the larger contribution  as $q_y$ is increased.  Our calculations show that the behaviors of the polarizability  when $\alpha=1.0$ are essentially the same as those for $\alpha=0.5$ in Figs.\  \ref{Fig2}(a) and \ref{Fig2}(b). However, we note that at large $q_y$ and high chemical potential,  the behavior of the plots is basically linear as it is for monolayer graphene.\cite{DD1,DD2,DD3,DD4,DD5}

\begin{figure}[h]
\centering
\includegraphics[width=1.0\textwidth]{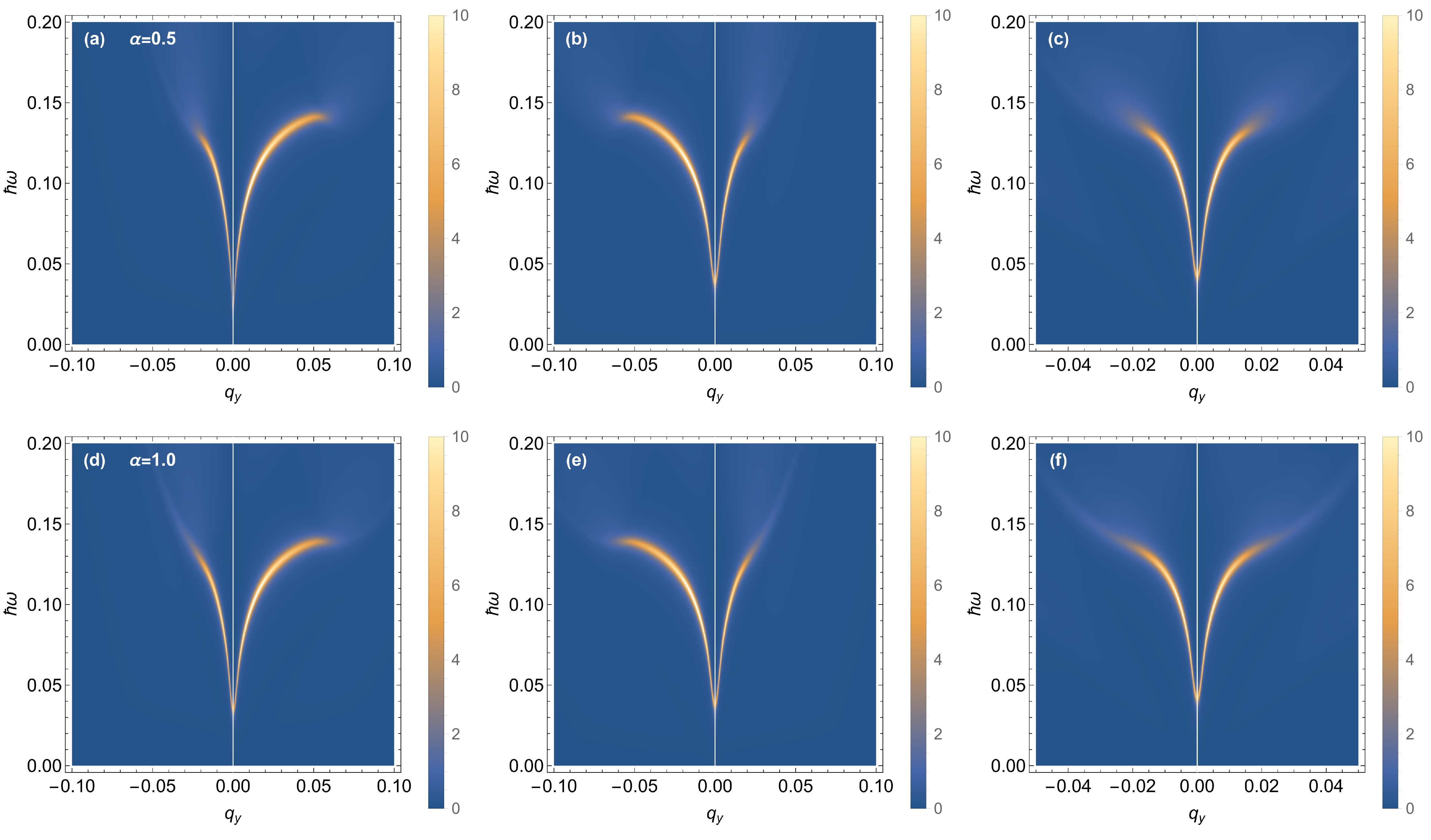}
\caption{ (Color online)   Plasmon dispersion along the $q_y$ direction when $\mu=0.1$. For the panels in the first row, $\alpha=0.5$ and $\lambda=-1$ in (a), $\lambda=1$ in (b) and the sum over both valleys $\lambda=\pm1$ in (c). In the second-row  panels, we chose $\alpha=1.0$ for the same values of $\lambda$ as the first row.}
\label{Fig3}
\end{figure}

\begin{figure}[h]
\centering
\includegraphics[width=1.0\textwidth]{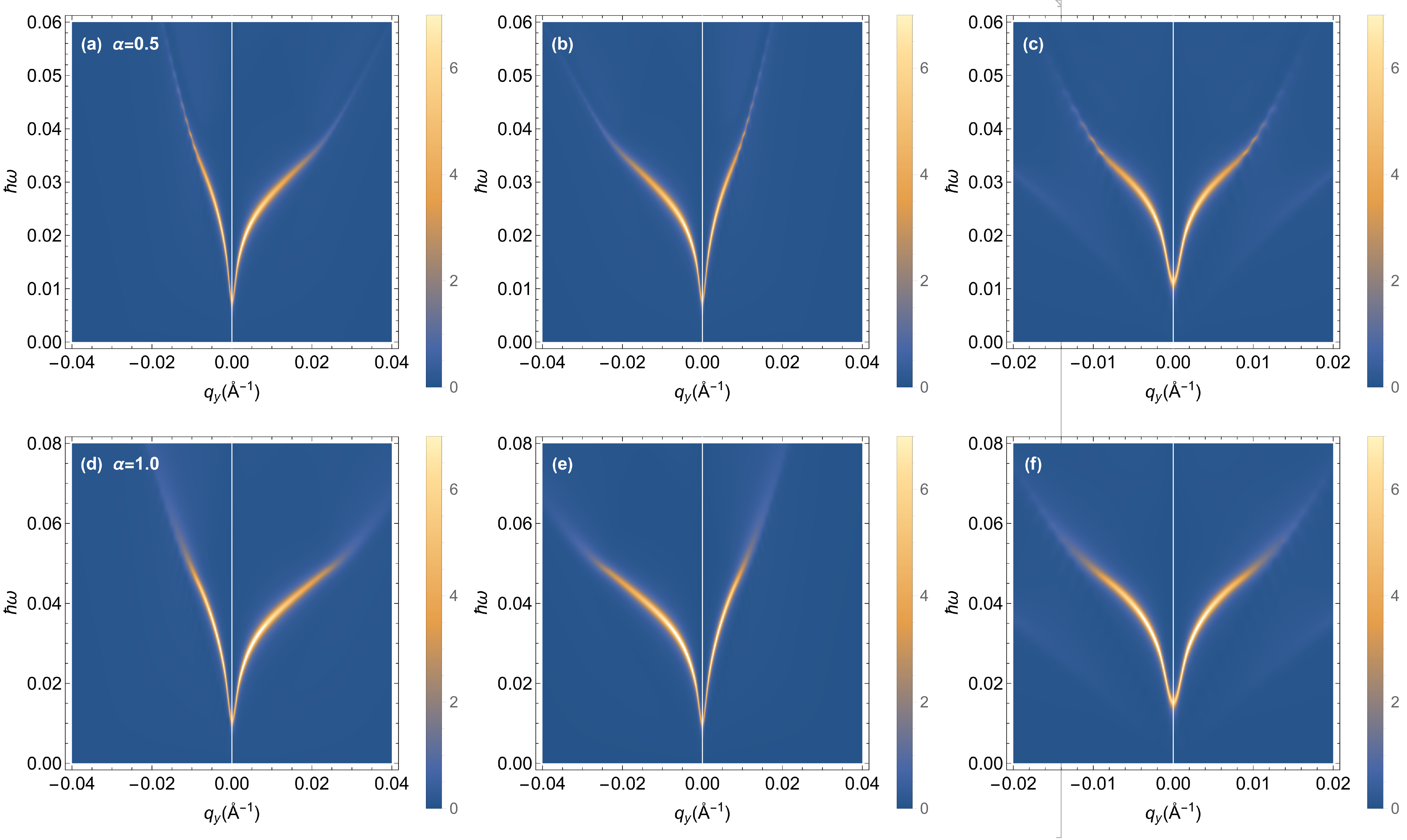}
\caption{ (Color online)   Plasmon dispersion along the $q_y$ direction. In the first row panels, $\alpha=0.5$, $\mu=0.03$ (just below the minimum of the higher conduction subband) and $\lambda=-1$ in (a), $\lambda=1$ in (b) and the sum over both valleys $\lambda=\pm1$ in (c). In the second row-panels we chose $\alpha=1.0$, $\mu=0.04$ (just below the minimum of the higher conduction subband) for the same values of $\lambda$ correspondingly as the first row.}
\label{Fig4}
\end{figure}

\section{Anisotropic Plasmon modes}
 \label{sec4}

 The energy dispersion for the self-sustained plasmon oscillations is determined by  the  zeros of the dielectric function, $\varepsilon({\bf q},\omega_p-i\gamma)=0$,  where $\varepsilon({\bf q},\omega)=1-v(q)\sum_\lambda \Pi_{\lambda}({\bf q},\omega)$, with $v(q)=e^2/2\varepsilon_0 q$ and the plasmon decay rate is
 
\begin{equation}
\gamma({\bf q},\omega_p)=-\  \frac{\text{Im}\  \Pi({\bf q},\omega_p)}{\partial \text{ Re}\  \Pi({\bf q},\omega)/\partial\omega_p} \ .
\label{gamma}
\end{equation}

\medskip
\par
In Fig.~\ref{Fig3},   we present the plasmon dispersion for $\alpha=0.5, \  1.0,\  \mu=0.1$ whereas in Fig.~ \ref{Fig4},   we present the plasmon dispersion for $\alpha=0.5,\  \mu=0.03$ and $\alpha=  1.0,\  \mu=0.04$. In the former case, the chemical potential crosses both conduction subbands whereas in the latter case, the chemical potential falls just below  the minimum of the  higher of the two conduction subbands. These results illustrate the anisotropy of the dispersion for both chosen values of $\alpha$, i.e., the plasma frequency depends on the direction of propagation. In all cases, the plasmon modes are Landau damped beyond a critical value of the wave vector $q_y$ which varies with $\mu$ and $\alpha$. As this Landau damping takes place within the particle-hole continuum,    we are principally interested in the plasmon branch in the region  outside this continuum, where $\text{Im}\Pi=0$ which yields  $\gamma=0$.  The presented results also illustrate that the two valleys contribute unequally. Comparing Figs.~\ref{Fig3} and   \ref{Fig4}, we see that the larger $\alpha$ has a greater group velocity in the long wavelength limit, thereby   accounting for  the collective properties of tilted MoS$_2$.

\medskip
\par
\begin{figure}[hbt!]
\centering
\includegraphics[width=1.0\textwidth]{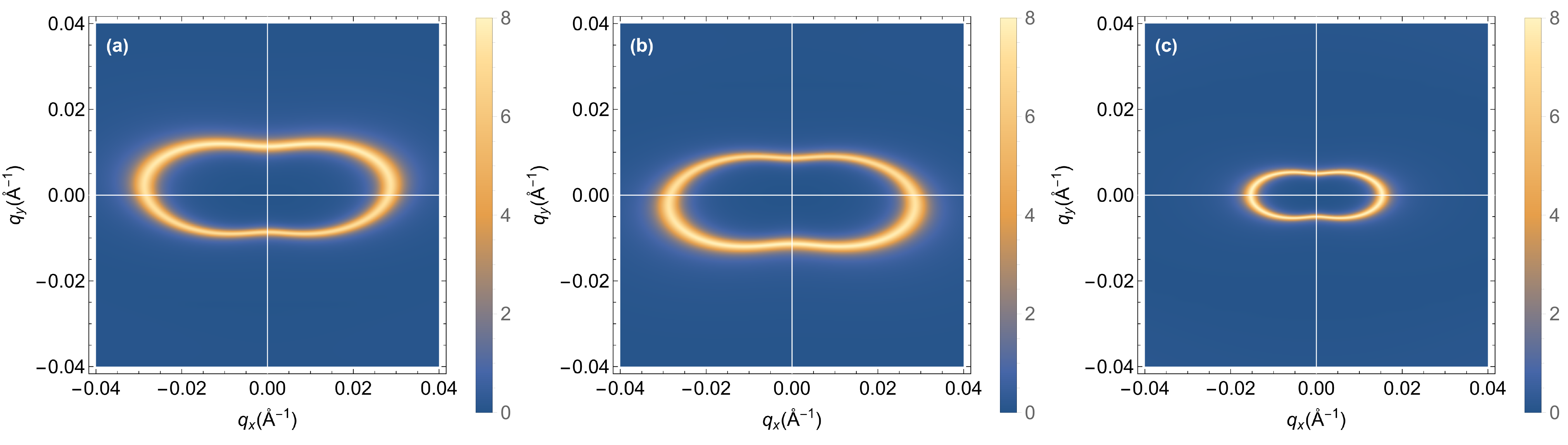}
\caption{ (Color online)   Plasmon isoenergy plot for $\alpha=0.5$, $\mu=0.1$ when (a) $\lambda=-1$, (b) $\lambda=1$ and (c) when we sum over both valleys $\lambda=\pm1$.  }
\label{Fig5}
\end{figure}

In Fig.~\ref{Fig5},  we present the plasmon dispersion for $\alpha=0.5, \  \mu=0.1$ where the plasmons isofrequency contours are shown when we sum over both valleys but intervalley terms are not included.    Both panels on the left and middle shown in Fig.~\ref{Fig5}  (a)  clearly illustrate the anisotropy of the plasmon dispersion is due to both valleys.  Figure\  \ref{Fig5} demonstrates that the breaking of the symmetry by the electric field leads to a dependence on the direction of propagation of the modes.  Choosing a frequency of oscillations in Fig.~\ref{Fig5} shows that this value can be achieved in  various directions of propagation.

\begin{figure}[ht]
\centering
\includegraphics[width=0.65\textwidth]{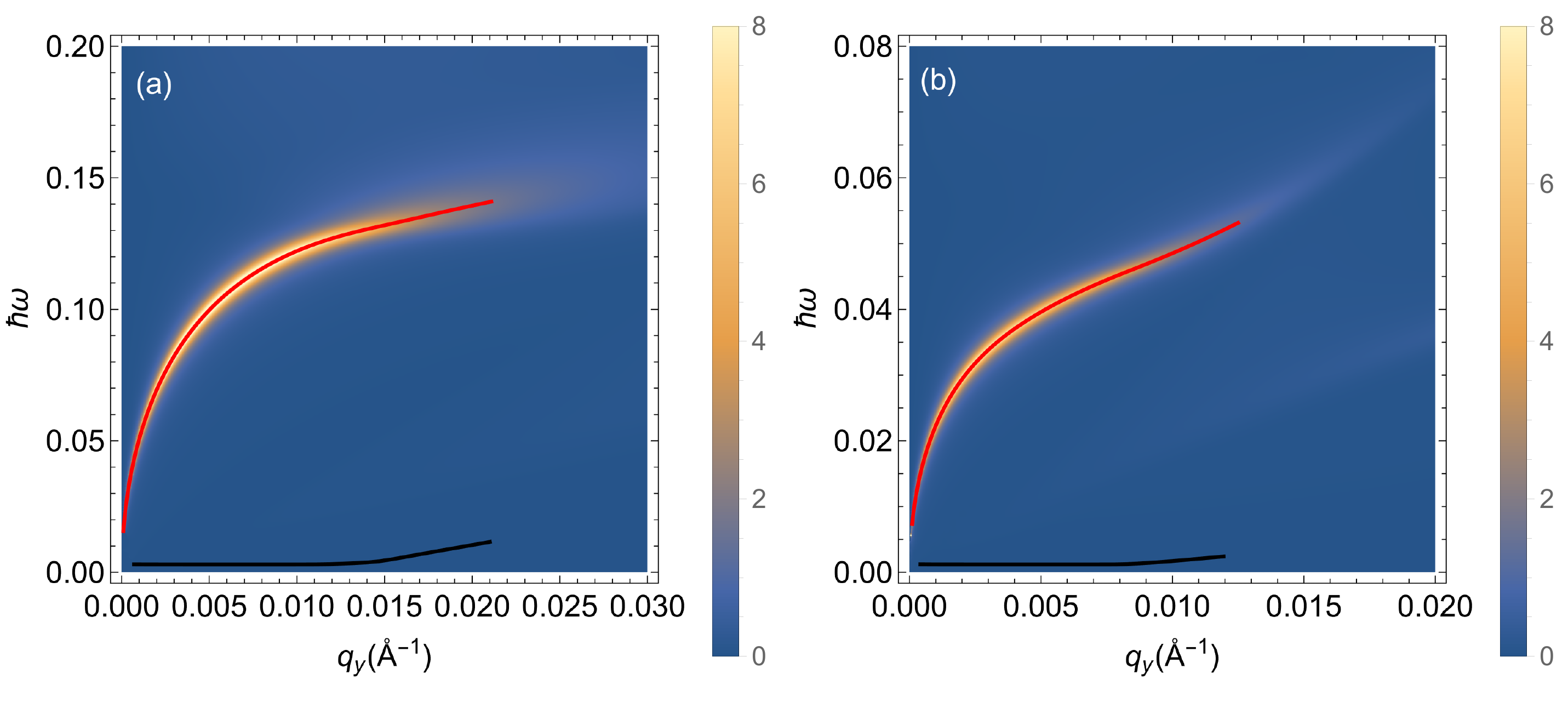}
\caption{ (Color online)  Decay rate $\gamma$ (lower black curve) of the plasmon (upper red curve)  versus $q_y$ for (a)\  $\alpha=0.5, \   \mu=0.1$  and (b)\  $\alpha=1.0, \   \mu=0.04$
when both valleys  contribute to the polarization.  In (a), the chemical potential is located in the upper conduction sub-band  (both  conduction sub-bands are partially filled).  In (b),  the chemical potential is located under the bottom of the  higher conduction sub-band (only the lowest conduction subband is occupied).
The unit for $\gamma$ is the same
as the energy unit used before for the plasmon dispersion, $\hbar q_y v_{-}$  with $q_y=1$\AA }.
\label{Fig9}
\end{figure}

Figure \ref{Fig9} shows the plasmon excitations and corresponding decay rates when $\alpha=0.5$ and $\alpha=1.0$. In  Fig.\  \ref{Fig9}(a), the chemical potential is $\mu=0.1$  located in the higher conduction subband so that both  conduction subbands are partially occupied.  However, in  Fig.\  \ref{Fig9}(b), the chemical potential is $\mu=0.04$  located below the higher conduction subband so that only one conduction sub-bands is partially occupied. The decay rate outside the particle hole region seems enhanced in Fig.~\ref{Fig9} shows the plasmon excitations and corresponding decay rates when $\alpha=0.5$. In  Fig.~\ref{Fig9} (a) for the higher  chemical potential thereby indicating the tunability of $\gamma$ for  1T$^\prime$-MoS$_2$.

 \section{ Static Screening Effects}
 \label{sec5}

We now employ the static limit of the polarizability $ \Pi({\bf q},\omega=0)$ or Lindhard function to calculate the screened Coulomb potential. This means that from the Lindhard function, it is possible to obtain the response of the Dirac fermions in the material to the presence of a magnetic or electric  impurity. The potential in the vicinity of a  point charge Q is proportional to the Fourier transform of the screened Coulomb interaction

\begin{equation}
\Phi({\bf r})=\frac{Q}{\epsilon_0}\int \frac{d^2{\bf q}}{(2\pi)^2}  \frac{v({\bf q})}{\epsilon({\bf q})} e^{i{\bf q}\cdot {\bf r}} \ ,
\label{SCR}
 \end{equation}
where $\epsilon({\bf q})= \epsilon({\bf q},\omega=0)$ and since the Lindhard function is anisotropic in ${\bf q}$-space, we must take into consideration its dependence on the polar angle of integration  in Eq.\ (\ref{SCR}).

\begin{figure}[hbt!]
\centering
\includegraphics[width=0.8\textwidth]{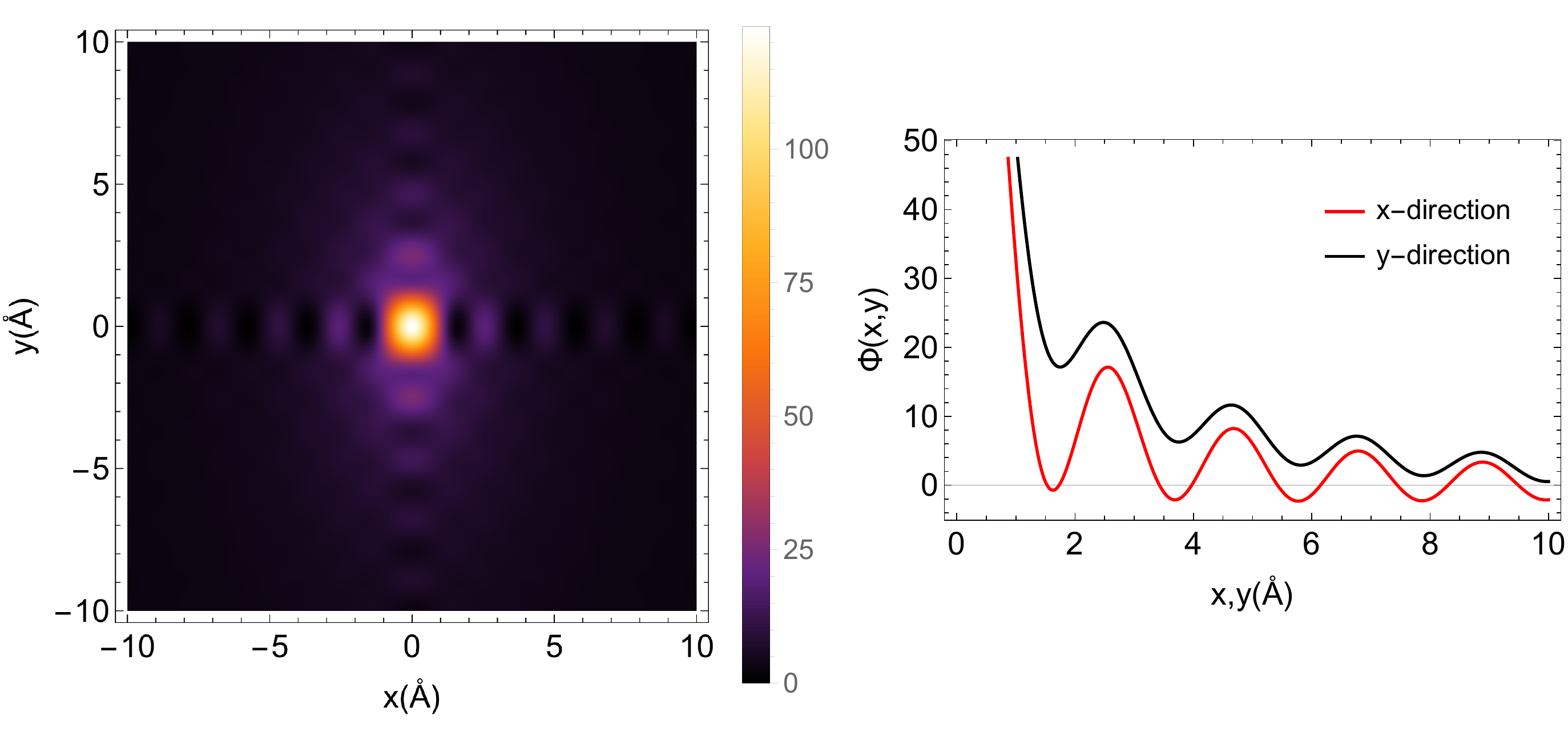}
\caption{ (Color online)   Screened potential $\Phi(x,y)$ for $\alpha=0.5$, $\mu=0.1$ with both valleys included in the dielectric function. In the left panel, we present a density plot for different distances $x,y$ from the screened charge $Q$. In the right panel, we see the variation of the potential in units of $48.1Q/\epsilon_0$\r{A} along the $x$ and $y$ directions. }
\label{Fig6}
\end{figure}

\medskip
\par
In Fig.~\ref{Fig6}, we present the static screened potential  for a dilute distribution of charge for which Eq.\  (\ref{SCR})  is applicable.  The calculated results show evidence of anisotropy along the mutually perpendicular $x$ and $y$  axes. The screened potential displays Friedel oscillations  with directional dependent amplitudes which are phase shifted. This is quite unlike the screened electrostatic potential for graphene  and the difference is attributed to the   significant variation  in their band structures.

 \section{Concluding Remarks}
 \label{sec6}

In conclusion, we have investigated the  behavior of the dynamical polarizability, plasmon excitations   as well as the static shielding of a dilute distribution of charged impurities  for 1T$^\prime$-MoS$_2$ in the presence of an external vertical electric field and strain. We also demonstrated that the plasmon damping rate generally increases for larger wavevector $q_y$ and larger frequencies along a plasmon branch. Therefore, we expect our damping rate to be monotonically increasing with $q_y$.  The tilted Dirac bands  which are valley-spin-polarized  cause   this material to undergo a topological phase change between a topological insulator and band insulator  corresponding to a critical value of the electric field.  We employed linear response theory   to calculate the polarization function which was  obtained numerically at T=0 K.   These results were then substituted into the dielectric function  for  calculating   the plasmon dispersion relation, their decay rate and impurity screening.
We would like to once again emphasize that here we have calculated both damped and undamped plasmons as zeros of the real part of the dielectric function.

\medskip
\par
Such distinctive features of 1T$^\prime$-MoS$_2$ are expected to give rise to a variety of applications   which  can be used for designing novel  multi-functional nanoelectronic and nanoplasmonic devices.  Specifically, the control of these collective properties could lead  to some important technological applications for electronic and optoelectronic devices.

\medskip
\par
It is evident that the electron dynamics in1T$^\prime$-MoS$_2$
under a vertical electric field and strain  is significantly different from that in graphene and this variation  can be controlled by an electrostatic potential.   This further implies that such a difference may be  tuned by a structure parameter $\alpha$   for 1T$^\prime$-MoS$_2$. 
These results presented here are expected to provide very useful information as well as guidance for designing nanoelectronic and nanoplasmonic devices
based on innovative low-dimensional 1T$^\prime$-MoS$_2$   materials

\medskip
\par

Finally, we observe that the presence of anisotropic plasmons in 1T$^\prime$-MoS$_2$ like
in 8-$Pmmn$ borophene   makes it possible for its use
for anisotropic plasma-wave photodetection, in a field-effect transistor–based device. 

\section*{Acknowledgment(s)}
G.G. would like to acknowledge the support from the Air Force Research Laboratory (AFRL) through Grant No. FA9453-21-1-0046
  
	\bibliographystyle{apsrev4-2}
	\bibliography{references}

\end{document}